\documentclass{llncs}
\usepackage[english]{babel}
\usepackage{amsmath,amssymb,amsfonts,stmaryrd}
\usepackage[all]{xypic}
\usepackage{pgf}
\usepackage{tikz}
\usepackage[utf8]{inputenc}
\usepackage{wasysym}
\newcommand{\com}[1]{}
\newcommand{\hiddencom}[1]{}
\newcommand{\manu}[1]{}
\def\A{{\mathbb{A}}}

\def\N{{\mathbb{N}}}

\def\R{{\mathbb{R}}}
\def\Z{{\mathbb{Z}}}
\def\Amaj{\cal A}

\def\Hmaj{\cal H}
\def\Imaj{\cal I}

\def\Pmaj{\cal P}
\def\ssbeq{\sqsubseteq}
\def\vect#1{\overrightarrow{#1}}
\def\idt{{\sf id}}
\newcommand{\dd}{\ensuremath{d}}
\title{A Geometric Approach\\ to the Problem of\\ Unique Decomposition of Processes}

\author{Thibaut Balabonski\inst{1} \and Emmanuel Haucourt\inst{2}}
\institute{Laboratoire PPS, Universit\'e Paris Diderot and CNRS, UMR 7126\\\email{thibaut.balabonski@pps.jussieu.fr}\\\  \and CEA, LIST, Gif-sur-Yvette, F-91191, France.\\\email{emmanuel.haucourt@cea.fr}}

\begin{document}

\maketitle\footnotetext{This work has been partially supported by {\em Agence Nationale pour la Recherche} via the project PANDA (Parallel and Distributed Analysis) ANR-09-BLAN-0169-02}

\begin{abstract}
  This paper proposes a geometric solution to the problem
  of prime decomposability of concurrent processes first explored by R. Milner and
  F. Moller in~\cite{RMFMDecomposition}. Concurrent programs are given
  a geometric semantics using cubical areas, for which a unique
  factorization theorem is proved. An effective factorization method
  which is correct and complete with respect to the geometric
  semantics is derived from the factorization theorem. This
  algorithm is implemented in the static analyzer \texttt{ALCOOL}.
\end{abstract}

\section{Introduction: Parallel Programming Problem}

This paper aims at introducing some new static analysis technology for
concurrent programs. The work presented here gives a new insight into
the problem of decomposition of processes, which was first explored by
{\em R. Milner} and {\em F. Moller} in~\cite{RMFMDecomposition}. The
main new results are an algorithm maximally decomposing concurrent
programs into independent processes (Section~\ref{algorithm}) and the
proof that this prime decomposition is unique in the considered class
of programs
(Theorem~\ref{monoid_of_cubical_areas_is_commutative_free}). They are
derived from a study of algebraic properties of cubical areas.

Given an associative and commutative operator $\|$ for parallel
composition of two processes (with the empty process as unit),
decomposing a concurrent program $P$ into a multiset
$\{P_1,...,P_n\}$ such that $P=P_1\|...\|P_n$ and the $P_i$s are
independent has several interests. For instance the decomposition may
be relevant for the allocation of processors to subprograms. Another
important concern is the static analysis of concurrent programs, whose
complexity grows exponentially with the number of concurrent
processes: finding independent subprograms that can be analyzed
separately could dramatically decrease the global complexity of the
static analysis. Hence this paper aims at finding the finest
decomposition (and proving its existence) for a wide class of concurrent
programs.

Let us first take a look at a non trivial example of independent
processes, in the so-called $PV$ language introduced by {\em
  E.W. Dijkstra}~\cite{EWDCooperating} as a simple framework for the
study of concurrency with shared resources. The only instructions are
$P(name)$ and $V(name)$\footnote{$P$ and $V$ stand for the dutch words
  ``Pakken'' (take) and ``Vrijlaten'' (release)}, where $name$ is an
identifier which refers to a resource.  The idea is to have some
common pool of resources which can be taken (with $P$) and released
(with $V$) by concurrent processes. The resources are formalized by
semaphores which, depending on their arity, can be held simultaneously
by a certain number of processes (arity $n$ allows at most $n-1$
simultaneous processes).

Now suppose $c$ is the name of a ternary semaphore, which means it can
be held by at most two processes, and $a,b$ are the names of binary
semaphores, also called {\em mutex} for {\em mutual exclusion}.
\begin{example}\label{IndExample}
\[\begin{array}{rcrcl}
\Sigma & := & \pi_1 & = & Pa.Pc.Vc.Va \\
& \| & \pi_2 & = & Pb.Pc.Vc.Vb \\
& \| & \pi_3 & = & Pa.Pc.Vc.Va \\
& \| & \pi_4 & = & Pb.Pc.Vc.Vb 
\end{array}
\]\end{example}
A naive syntactic analysis would stamp this program as undecomposable
since all processes share the resource $c$, but the following finer
analysis can be made: thanks to mutex $a$ (respectively $b$), the
processes $\pi_1$ and $\pi_3$ (respectively $\pi_2$ and $\pi_4$)
cannot both hold an occurrence of the resource $c$ at the same
time\manu{at any given time at most one among $\pi_1$ and $\pi_3$
  (respectively $\pi_2$ and $\pi_4$) tries to use the resource
  $c$}\manu{, and the same holds for $\pi_2$ and $\pi_4$ with mutex
  $b$}. Then there are never more than two simultaneous requests over
$c$, which means that the instructions $Pc$ and $Vc$ play actually no
role in determining the semantics of the program. And without $c$,
$\Sigma$ can be split in two independent systems (they use disjoint
resources).  Basically, this example is based on the fact that
semaphores are not the real resources, but mere devices used to guard
their access. And it may be that some guards are redundant.

This work is based on a geometric semantics for concurrency. The
semantics for $PV$ programs was implicitly given
in~\cite{EWDCooperating}, then explicited by {\em Carson} et
al.\cite{SDCPFRGeometry}. Roughly speaking, the instructions of a
process are pinned upon a $1$-dimensional ``directed'' shape, in other
words track along which the instructions of the program to execute are
written. If $N$ sequential processes run together, one can consider
their $N$ instruction pointers as a multi-dimensional control point.

Although we have made the construction explicit for $PV$ programs
only, the result applies to any synchronisation or communication
mechanism whose geometric interpretation is a so-called {\em cubical
  area} (the notion is formalized in Section~\ref{Cubical_Areas}). See
for instance~\cite{EGEHPraAppGeo} for the geometric semantics of
synchronisation barriers, monitors and synchronous or asynchronous
communications (with finite or infinite message queues): their
geometrical shape is the complement of an orthogonal
polyhedron~\cite{BMPOrtho,DTHabilitation}, which is a special case of
cubical area.

\medskip\noindent{\bf Outline of the paper.}\\ The paper is organized
as follows. Section 2 provides the mathematics of the geometric
semantics, detailed for $PV$ programs. Section 3
establishes the link between algebraic properties of the semantics and
independence of subprograms, and then states and proves prime decomposability
theorems for algrebraic frameworks encompassing the geometric
semantics (Theorems~\ref{homogeneous_is_commutative_free} and~
\ref{monoid_of_cubical_areas_is_commutative_free}).
Section 4 describes the corresponding algorithm and
implementation as well as a detailed example and some benchmarks.

\section{The Geometric Semantics}
%
The geometric semantics of a PV program is a subset of the finite
dimensional real vector space whose dimension is roughly speaking the
number $N$ of processes running concurrently. Then each process is
associated with a coordinate of $\R^N$. Yet given a mutex \texttt{a},
the instructions \texttt{P(a)} and \texttt{V(a)} that occur in the
$k^{\text{th}}$ process should be understood as opening and closing
parentheses or more geometrically as the least upper bound and the 
greatest lower bound of an interval $I_k$ of $\R$. The forbidden area
generated by a mutex \texttt{a} is thus the finite union of 
hyperrectangles\footnote{however we will more
  likely write ``cube'' instead.} of the following form (with $k<k'$)
$$
\underbrace
{\R^+\times\cdots\times\R^+\times I_k\times\R^+\times\cdots\times\R^+\times I_{k'}\times\R^+\times\cdots\times\R^+}_{\text{product of $N$ terms}}
$$
\begin{minipage}{0.6\linewidth}
For example, \texttt{P(a).V(a) $\|$ P(a).V(a)} is a program written in PV
language. Assuming that $a$ is a mutex (semaphore of arity $2$), its
geometric model is $(\R^+)^2\backslash[1,2[^2$. Intuitively, a point $p$
    in $[1,2[^2$ would correspond to the situation where both
        processes hold the semaphore $a$, which is forbidden by the
        semantics of mutices. 
\end{minipage}
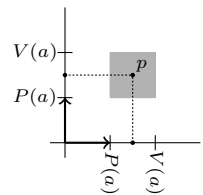
\begin{minipage}{0.38\linewidth}
\begin{center}
\begin{tikzpicture}
\begin{scope}
\draw[fill,black!30] (6mm,6mm) rectangle (12mm,12mm) ;
\draw[->,thick] (0mm,0mm) -- (6mm,0mm) ;
\draw[->,thick] (0mm,0mm) -- (0mm,6mm) ;
\fill (9mm,9mm) circle (0.3mm) ;
\draw (10.2mm,10.2mm) node{$\scriptstyle p$} ;
\draw[-] (-2mm,0mm) -- (18mm,0mm) ;
\draw[-] (0mm,-2mm) -- (0mm,18mm) ;
\draw[thin] (6mm,-1mm) -- (6mm,1mm) ;
\draw[thin] (12mm,-1mm) -- (12mm,1mm) ;
\draw[thin] (-1mm,6mm) -- (1mm,6mm) ;
\draw[thin] (-1mm,12mm) -- (1mm,12mm) ;
\draw[thin,dash pattern = on 0.25mm off 0.25mm] (0mm,9mm) -- (9mm,9mm) ;
\draw[thin,dash pattern = on 0.25mm off 0.25mm] (9mm,0mm) -- (9mm,9mm) ;
\draw (6mm,-4mm) node[rotate = -90]{$\scriptstyle P(a)$} ;
\draw (12mm,-4mm) node[rotate = -90]{$\scriptstyle V(a)$} ;
\draw (-4mm,6mm) node{$\scriptstyle P(a)$} ;
\draw (-4mm,12mm) node{$\scriptstyle V(a)$} ;
\fill (0mm,9mm) circle (0.3mm) ;
\fill (9mm,0mm) circle (0.3mm) ;
\end{scope}
\end{tikzpicture}
\end{center}
\end{minipage}
\\\ \\
In the sequel of this section we formalize the PV language syntax as
well as the construction of the geometric semantics.
Denote the positive half-line $[0,+\infty[$ by $\R^+$. For each
    $\alpha\in\N\backslash\{0,1\}$ let $S_\alpha$ be an infinite
    countable set whose elements are the semaphores of arity $\alpha$
    of the PV language. A {\bf PV process} is a finite sequence on the
    alphabet
$$
A:=\big\{P(s),V(s)\ \big|\ s\in\hspace{-1mm}\bigcup_{\alpha\geq 2}\hspace{-1mm}S_\alpha\big\}
$$ and a {\bf PV program} is a finite (and possibly empty) multiset of
PV processes. The parallel operator then corresponds to the multiset
addition therefore it is associative and commutative \footnote{The
  collection of multisets over a set $\A$ forms a monoid which is
  isomorphic to the free commutative monoid over $\A$. The first
  terminology is usually used by computer scientists while
  mathematicians prefer the second one. Anyway it will be described
  and caracterized in the Section 3.}. Given a semaphore $s$ and a
process $\pi$, the sequences $(x_k)_{k\in\N}$ and $(y_k)_{k\in\N}$ are
recursively defined as follows: set $y_{-1}=0$ and
\begin{itemize}
\item $x_k=\min\{n\in\N\ |\ n> y_{k-1}\ \mbox{and}\
\pi(n)\ \mbox{is}\ P(s)\}$
\item $y_k=\min\{n\in\N\ |\ n> x_k\ \mbox{and}\
\pi(n)\ \mbox{is}\ V(s)\}$
\end{itemize}
with the convention that $\min\emptyset=\infty$, $\pi(n)$ denotes the
$n^{\mbox{\tiny th}}$ term of the process $\pi$ and its first term is
$\pi(1)$. Then, the {\bf busy area} of $s$ in $\pi$
is\footnote{Including the greatest lower bound and removing the least
upper one is the mathematical interpretation of the following
convention: the changes induced by an instruction are effective
exactly when the instruction pointer reaches it.}
$$
B_s(\pi):=\bigcup_{k\in\N}[x_k,y_k[
$$ Actually this description requires some extra assumptions upon the
    way instructions are interpreted. Namely a process cannot hold
    more than one occurrence of a given ressource. Thus a process
    already holding an occurrence of a semaphore $s$ ignores any
    instruction $P(s)$, and similarly a process holding no occurrence
    of $s$ ignores any instruction $V(s)$.  \hiddencom{Ces définitions
      imposent une certaine sémantique au langage, qu'il faudrait
      préciser: chaque processus ne peut tenir qu'une occurrence de
      chaque ressource, et les opérations ``interdites'' sont ignorées
      (P quand on possède déjà une occurrence de la ressource ou V
      quand ce n'est pas le cas).} Then denote by
    $\chi_s^{\pi}:\R\rightarrow\R$ the characteristic function of
    $B_s$ defined by
$$
\chi_s^{\pi}(x)=
\left\{
\begin{array}{ll}
1 & \mbox{if}\ x\in B_s(\pi) \\
0 & \mbox{otherwise}
\end{array}
\right.
$$
\noindent
Because the sequence $\pi$ is finite, there exists some
$k$ such that $x_k=\infty$ and for any such $k$ and any $k'\geq k$,
one also has $x_{k'}=\infty$. In particular, if the instruction $P(s)$
does not appear in $\pi$, then $B_s(\pi)$ is empty and $\chi_s^{\pi}$
is the null map.  The geometric model of a PV program with $N$
processes running concurrently is a subpospace of $[0,+\infty[^N$
defined as follows:
\\
- Call $\Pi=(\pi_1,\ldots,\pi_N)$ the program to modelize.
\\
- Given a semaphore $s$ of arity $\alpha$ define the {\bf forbidden
area} of $s$ in $\Pi$ as
$$
F_s:=\big\{\vect{x}\in[0,+\infty[^N\ \big|\ \vect{\chi_s}\cdot\vect{x}\geq\alpha \big\}
$$ where $\vect{x}=(x_1,\ldots,x_N)$,
    $\vect{\chi_s}=(\chi_s^{\pi_1},\ldots,\chi_s^{\pi_N})$ and $
    \vect{\chi_s}\cdot\vect{x}=\overset{N}{\underset{i=1}{\sum}}\chi_s^{\pi_i}(x_i)
    $.  The value $\vect{\chi_s}\cdot\vect{x}$ indicates how many
    occurrences of the semaphore $s$ are held when the instruction
    pointer is at position $\vect{x}$. Note that $F_s$ is a finite
    union of hyperrectangles which may be empty even if $s$ appears in
    the program $\Pi$. In the end, the {\bf forbidden area} of the
    program $\Pi$ is the following union over $S$ the union of all the
    sets $S_\alpha$.
$$
F:=\bigcup_{s\in S} F_s
$$
Because there are finitely many resource names $s$ appearing in
a PV program, there are finitely many non empty set $F_s$. Hence the
previous union is still a finite union of hyperrectangles. The {\bf
  state space} or {\bf geometric model} of $\Pi$ is then
$[0,+\infty[^N\backslash F$, and is denoted by
    $\llbracket\Pi\rrbracket$. Remark that the geometric model is also
    a finite union of hyperrectangles.
    
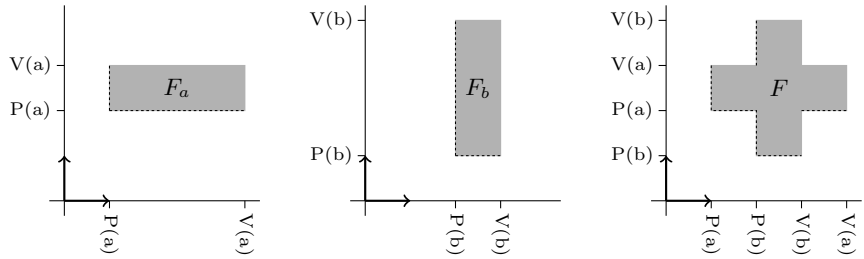
\begin{figure}
\caption{Construction of a geometric model: the swiss flag\label{SwissFlag}}
\begin{center}
\begin{tikzpicture}
\begin{scope}
\draw[fill,black!30] (6mm,12mm) rectangle (24mm,18mm) ;
\draw[dash pattern = on 0.35mm off 0.35mm] (6mm,12mm) -- (6mm,18mm) ;
\draw[dash pattern = on 0.35mm off 0.35mm] (6mm,12mm) -- (24mm,12mm) ;
\draw[->,thick] (0mm,0mm) -- (6mm,0mm) ;
\draw[->,thick] (0mm,0mm) -- (0mm,6mm) ;
\draw[-] (-2mm,0mm) -- (26mm,0mm) ;
\draw[-] (0mm,-2mm) -- (0mm,26mm) ;
\draw (15mm,15mm) node{$F_a$} ;
\draw[-] (6mm,-1mm) -- (6mm,0mm) ;
\draw[-] (24mm,-1mm) -- (24mm,0mm) ;
%
\draw[-] (-1mm,12mm) -- (0mm,12mm) ;
\draw[-] (-1mm,18mm) -- (0mm,18mm) ;
%
\draw (-4.5mm,12mm) node{\scriptsize P(a)} ;
\draw (-4.5mm,18mm) node{\scriptsize V(a)} ;
\draw (6mm,-4.5mm)  node[rotate = -90]{\scriptsize P(a)} ;
\draw (24mm,-4.5mm) node[rotate = -90]{\scriptsize V(a)} ;
\end{scope}
\begin{scope}[xshift = 40mm]
\draw[fill,black!30] (12mm,6mm) rectangle (18mm,24mm) ;
\draw[dash pattern = on 0.35mm off 0.35mm] (12mm,6mm) -- (18mm,6mm) ;
\draw[dash pattern = on 0.35mm off 0.35mm] (12mm,6mm) -- (12mm,24mm) ;
\draw (15mm,15mm) node{$F_b$} ;
\draw[->,thick] (0mm,0mm) -- (6mm,0mm) ;
\draw[->,thick] (0mm,0mm) -- (0mm,6mm) ;
\draw[-] (-2mm,0mm) -- (26mm,0mm) ;
\draw[-] (0mm,-2mm) -- (0mm,26mm) ;
%
\draw[-] (12mm,-1mm) -- (12mm,0mm) ;
\draw[-] (18mm,-1mm) -- (18mm,0mm) ;
%
\draw[-] (-1mm,6mm) -- (0mm,6mm) ;
\draw[-] (-1mm,24mm) -- (0mm,24mm) ;
\draw (-4.5mm,6mm)  node{\scriptsize P(b)} ;
\draw (-4.5mm,24mm) node{\scriptsize V(b)} ;
%
\draw (12mm,-4.5mm) node[rotate = -90]{\scriptsize P(b)} ;
\draw (18mm,-4.5mm) node[rotate = -90]{\scriptsize V(b)} ;
\end{scope}
\begin{scope}[xshift = 80mm]
\draw[fill,black!30] (6mm,12mm) rectangle (24mm,18mm) ;
\draw[fill,black!30] (12mm,6mm) rectangle (18mm,24mm) ;
\draw[dash pattern = on 0.35mm off 0.35mm] (6mm,18mm) -- (6mm,12mm) -- (12mm,12mm) -- (12mm,6mm)  -- (18mm,6mm);
\draw[dash pattern = on 0.35mm off 0.35mm] (12mm,18mm) -- (12mm,24mm) ;
\draw[dash pattern = on 0.35mm off 0.35mm] (18mm,12mm) -- (24mm,12mm) ;
\draw (15mm,15mm) node{$F$} ;
\draw[->,thick] (0mm,0mm) -- (6mm,0mm) ;
\draw[->,thick] (0mm,0mm) -- (0mm,6mm) ;
\draw[-] (-2mm,0mm) -- (26mm,0mm) ;
\draw[-] (0mm,-2mm) -- (0mm,26mm) ;
\draw[-] (6mm,-1mm) -- (6mm,0mm) ;
\draw[-] (12mm,-1mm) -- (12mm,0mm) ;
\draw[-] (18mm,-1mm) -- (18mm,0mm) ;
\draw[-] (24mm,-1mm) -- (24mm,0mm) ;
\draw[-] (-1mm,6mm) -- (0mm,6mm) ;
\draw[-] (-1mm,12mm) -- (0mm,12mm) ;
\draw[-] (-1mm,18mm) -- (0mm,18mm) ;
\draw[-] (-1mm,24mm) -- (0mm,24mm) ;
\draw (-4.5mm,6mm)  node{\scriptsize P(b)} ;
\draw (-4.5mm,12mm) node{\scriptsize P(a)} ;
\draw (-4.5mm,18mm) node{\scriptsize V(a)} ;
\draw (-4.5mm,24mm) node{\scriptsize V(b)} ;
\draw (6mm,-4.5mm)  node[rotate = -90]{\scriptsize P(a)} ;
\draw (12mm,-4.5mm) node[rotate = -90]{\scriptsize P(b)} ;
\draw (18mm,-4.5mm) node[rotate = -90]{\scriptsize V(b)} ;
\draw (24mm,-4.5mm) node[rotate = -90]{\scriptsize V(a)} ;
\end{scope}
\end{tikzpicture}
\end{center}
\end{figure}

\begin{figure}
\caption{\label{3DExample} Example in three dimensions}
\begin{center}
\begin{tabular}{c@{\hspace{0.7cm}}c@{\hspace{0.7cm}}c}
\begin{tikzpicture}
\fill (0.5,0.3,-2.5) circle (3pt);
\draw[thick,->] (0.5,0.3,-0.7) -- (0.5,0.3,-2.3);
\draw[thick,->] (0.6,0.3,-2.5) -- (2.4,0.3,-2.5);
\fill (0.5,1.7,-2.5) circle (3pt);
\draw[thick,->] (0.5,1.7,-0.7) -- (0.5,1.7,-2.3);
\draw[thick,->] (0.6,1.7,-2.5) -- (2.4,1.7,-2.5);
\draw[thick,->] (0.5,0.4,-2.5) -- (0.5,1.6,-2.5);
\fill[color=black!45!white, opacity=0.9] (1,-0.4,-1) -- (2,-0.4,-1) -- (2,-0.4,-2) -- (1,-0.4,-2) -- cycle;
\fill[color=black!45!white, opacity=0.9] (1,-0.4,-2) -- (2,-0.4,-2) -- (2,2.3,-2) -- (1,2.3,-2) -- cycle;
\fill[color=black!45!white, opacity=0.9] (1,-0.4,-1) -- (1,-0.4,-2) -- (1,2.3,-2) -- (1,2.3,-1) -- cycle;
\draw[color=white!80!black] (1,-0.4,-1) -- (1,-0.4,-2) -- (2,-0.4,-2) (1,-0.4,-2) -- (1,2.3,-2);
\fill (1.2,1.15,-1.2) -- (1.8,1.15,-1.2) -- (1.8,1.15,-1.8) -- (1.2,1.15,-1.8) -- cycle;
\fill (1.2,0.55,-1.2) -- (1.8,0.55,-1.2) -- (1.8,1.15,-1.2) -- (1.2,1.15,-1.2) -- cycle;
\fill (1.8,0.55,-1.2) -- (1.8,0.55,-1.8) -- (1.8,1.15,-1.8) -- (1.8,1.15,-1.2) -- cycle;
\fill[color=black!50!white, opacity=0.7] (1,2.3,-1) -- (2,2.3,-1) -- (2,2.3,-2) -- (1,2.3,-2) -- cycle;
\fill[color=black!50!white, opacity=0.7] (1,-0.4,-1) -- (2,-0.4,-1) -- (2,2.3,-1) -- (1,2.3,-1) -- cycle;
\fill[color=black!50!white, opacity=0.7] (2,-0.4,-1) -- (2,-0.4,-2) -- (2,2.3,-2) -- (2,2.3,-1) -- cycle;
\draw (1.8,1.15,-1.8) -- (1.2,1.15,-1.8) -- (1.2,1.15,-1.2) -- (1.2,0.55,-1.2) -- (1.8,0.55,-1.2) -- (1.8,0.55,-1.8) -- (1.8,1.15,-1.8) -- (1.8,1.15,-1.2) -- (1.2,1.15,-1.2) (1.8,1.15,-1.2) -- (1.8,0.55,-1.2);
\draw[very thin,color=black!80!white] (1.2,0.55,-1.2) -- (1.2,0.55,-1.8) -- (1.8,0.55,-1.8) (1.2,0.55,-1.8) -- (1.2,1.15,-1.8);
\draw[color=white!70!black] (1,2.3,-1) -- (2,2.3,-1) -- (2,2.3,-2) (2,2.3,-1) -- (2,-0.4,-1);
\fill (0.5,0.3,-0.5) circle (3pt);
\fill (2.5,0.3,-0.5) circle (3pt);
\fill (2.5,0.3,-2.5) circle (3pt);
\draw[thick,->] (2.5,0.3,-0.7) -- (2.5,0.3,-2.3);
\draw[thick,->] (0.6,0.3,-0.5) -- (2.4,0.3,-0.5);
\fill (0.5,1.7,-0.5) circle (3pt);
\fill (2.5,1.7,-0.5) circle (3pt);
\fill (2.5,1.7,-2.5) circle (3pt);
\draw[thick,->] (2.5,1.7,-0.7) -- (2.5,1.7,-2.3);
\draw[thick,->] (0.6,1.7,-0.5) -- (2.4,1.7,-0.5);
\draw[thick,->] (0.5,0.4,-0.5) -- (0.5,1.6,-0.5);
\draw[thick,->] (2.5,0.4,-0.5) -- (2.5,1.6,-0.5);
\draw[thick,->] (2.5,0.4,-2.5) -- (2.5,1.6,-2.5);
\end{tikzpicture}
&
\raisebox{1.4cm}{
$
\begin{array}{rcrcl}
\Sigma^* := & \pi_1 & = & Pa.Pc.Vc.Va \\
\| & \pi^*_2 & = & \phantom{Pb.}Pc.Vc\phantom{.Vb} \\
\| & \pi_3 & = & Pa.Pc.Vc.Va
\end{array}
$}
&
\begin{tikzpicture}
\fill (0.5,1,-2.5) circle (3pt);
\draw[thick,->] (0.5,1,-1) -- (0.5,1,-2.3);
\draw[thick,->] (0.6,1,-2.5) -- (2.4,1,-2.5);
\fill[color=black!45!white, opacity=0.9] (1,-0,-1) -- (2,-0,-1) -- (2,-0,-2) -- (1,-0,-2) -- cycle;
\fill[color=black!45!white, opacity=0.9] (1,-0,-2) -- (2,-0,-2) -- (2,2,-2) -- (1,2,-2) -- cycle;
\fill[color=black!45!white, opacity=0.9] (1,-0,-1) -- (1,-0,-2) -- (1,2,-2) -- (1,2,-1) -- cycle;
\draw[color=white!80!black] (1,-0,-1) -- (1,-0,-2) -- (2,-0,-2) (1,-0,-2) -- (1,2,-2);
\fill[color=black!50!white, opacity=0.7] (1,2,-1) -- (2,2,-1) -- (2,2,-2) -- (1,2,-2) -- cycle;
\fill[color=black!50!white, opacity=0.7] (1,-0,-1) -- (2,-0,-1) -- (2,2,-1) -- (1,2,-1) -- cycle;
\fill[color=black!50!white, opacity=0.7] (2,-0,-1) -- (2,-0,-2) -- (2,2,-2) -- (2,2,-1) -- cycle;
\draw[color=white!70!black] (1,2,-1) -- (2,2,-1) -- (2,2,-2) (2,2,-1) -- (2,-0,-1);
\fill (2.5,1,-2.5) circle (3pt);
\fill[color = white!75!black,opacity = 0.7] (1,-0,-1) -- (0,-0,-1) -- (0,2,-1) -- (1,2,-1)-- cycle;
\draw[very thin] (1,-0,-1) -- (0,-0,-1) -- (0,2,-1) -- (1,2,-1)-- cycle;
\fill (0.5,1,-0.5) circle (3pt);
\fill[color = white!75!black,opacity = 0.7] (2,-0,-2) -- (2,-0,-3) -- (2,2,-3) -- (2,2,-2) -- cycle;
\draw[very thin] (2,-0,-2) -- (2,-0,-3) -- (2,2,-3) -- (2,2,-2) -- cycle;
\draw[thick,->] (2.5,1,-1.9) -- (2.5,1,-2.3);
\draw[thick] (0.6,1,-0.5) -- (1,1,-0.5);
\fill[color = white!75!black,opacity = 0.7] (2,-0,-2) -- (3,-0,-2) -- (3,2,-2) -- (2,2,-2)-- cycle;
\fill[color = white!75!black,opacity = 0.7] (1,-0,-1) -- (1,-0,0) -- (1,2,0) -- (1,2,-1) -- cycle;
\draw[very thin] (2,-0,-2) -- (3,-0,-2) -- (3,2,-2) -- (2,2,-2)-- cycle;
\draw[very thin] (1,-0,-1) -- (1,-0,0) -- (1,2,0) -- (1,2,-1) -- cycle;
\draw[thick,->] (1,1,-0.5) -- (2.4,1,-0.5);
\draw[thick] (0.5,1,-0.7) -- (0.5,1,-1.1);
\draw[thick] (2.5,1,-0.7) -- (2.5,1,-1.9);
\fill (2.5,1,-0.5) circle (3pt);
\end{tikzpicture}
\end{tabular}
\end{center}
Intuitively, the graphs pictured here correspond to the {\em essential
  components} of the state space, see~\cite{EGEHCompCat2} for
developments on this topic.
The dark grey cube on the left picture is the forbidden area of the
semaphore $c$, which is contained in the forbidden area of the mutex
$a$ (in the full --and 4D-- example $\Sigma$ the forbidden area of $c$
is contained in the union of the forbidden areas of $a$ and $b$).
\end{figure}
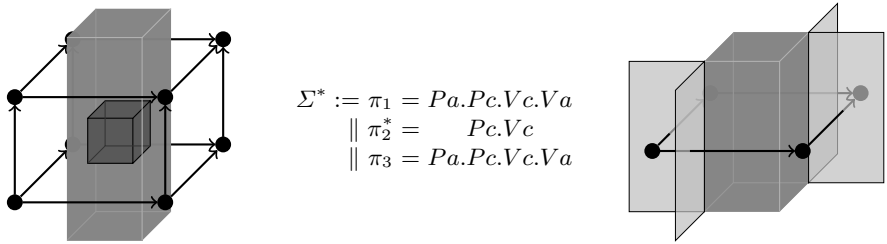

In other words, the state space of $\Pi$ is the set of positions of
the ``multi dimensional instruction pointer'' for which the number of
occurrences of each semaphore $s$ is strictly below its arity
$\alpha$.  If $\Pi$ is made of $N$ concurrent process, this space is a
$N$-dimensional enclidean space with (cubical) holes.  As an example,
Figure~\ref{SwissFlag} shows the construction of the geometric model
of the PV program $P(a)P(b)V(b)V(a)\ \|\ P(b)P(a)V(a)V(b)$ (refered to
as the {\em swiss flag}). Figure~\ref{3DExample} gives a simplified
version of Example~\ref{IndExample} fitting in three dimensions.
\com{Ce dessin montrait la catégorie des composantes, qui n'existe
  plus ici, je devrais le modifier un peu pour ne pas perturber le
  lecteur.}

%

%

\section{The Problem of Unique Decomposition}
Now that the geometric semantics of programs is defined, let us
refocus on the main goal: finding the independent parts of a concurrent program.
Hence the question: what does independence mean in this geometrical setting?
%
\subsection{Parallel Composition vs Cartesian Product}
A general definition has to be given for {\bf independence}: say a
program $\Pi$ is independent from another program $\Pi'$ when its
behaviour is unaffected by parallel composition with $\Pi'$, whatever
the way $\Pi'$ is executed. That means, the presence of $\Pi'$, as
well as its instruction pointer, has no effect on the semantics of
$\Pi$. A geometric translation of this assertion is: in the
geometric model of $\Pi \| \Pi'$, the cylinder\footnote{Categorists
  would write ``fibre'' instead of ``cylinder''.} over any state of
$\Pi'$ (i.e.\ the subspace of all points with given fix coordinates
for the $\Pi'$ component) is equal to the model of
$\Pi$.\hiddencom{J'ai peur que le terme ``fibre'' ne soit pas compris
  par les non-catégoriciens, qui sont probablement nombreux à concur,
  alors que ``cylindre'' reprend de la géométrie élémentaire.}
\hiddencom{C'est la raison pour laquelle j'ai précisé la définition
  entre parenthèse. Ceci étant dit c'est peut-ếtre plus sage.}

Hence two programs $\Pi$ and $\Pi'$ of geometric models
$\llbracket\Pi\rrbracket$ and $\llbracket\Pi'\rrbracket$ are
indenpendent if and only if the geometric model
$\llbracket\Pi\|\Pi'\rrbracket$ of their parallel composition is
isomorphic to the cartesian product
$\llbracket\Pi\rrbracket\times\llbracket\Pi'\rrbracket$.  Thus the
decompositions of a program correspond to the factorizations if its
geometric model (with respect to the cartesian product).
Next subsection reminds some algebraic settings and results needed for
a notion like {\em factorization} to make sense.

\manu{\noindent
The decomposition of orthogonal polyhedra defined in
\cite{DTHabilitation} is understood with respect to the binary set
union instead of Cartesian product, so it has nothing to do with our
notion though the similar terminology might be confusing.}

\manu{\noindent
The sequel of the section is devoted to the theoretical cornerstone of
the article.}

\subsection{Free Commutative Monoids}
\hiddencom{Il faudrait un renvoi à un manuel d'algèbre de référence pour
  appuyer toute cette section.}
The reader not familiar with this notion can refer for instance to
\cite{Algebra}.
Let $M$ be a commutative monoid. Any element of $M$ which has an
inverse is called a {\bf unit}. A {\em non unit} element $x$ of $M$ is
said to be {\bf irreducible} when for all $y$ and $z$ in $M$, if
$x=yz$ then $y$ or $z$ is a unit. The set of irreducible elements of
$M$ is denoted by $I(M)$.

For any elements $x$ and $y$ of $M$, say $x$ {\bf divides} $y$ when
there is an element $x'$ of $M$ such that $xx'=y$.  A {\em non unit}
element $x$ of $M$ is said to be {\bf prime} when for all $y$ and $z$
in $M$, if $x$ divides $yz$ then $x$ divides $y$ or $x$ divides
$z$. The set of prime elements of $M$ is denoted by $P(M)$.

Given a set $X$, the collection of maps $\phi$ from $X$ to $\N$ such
that $\{x\in X\ |\ \phi(x)\not=0\}$ is finite, together with the
pointwise addition, forms a commutative monoid whose neutral element
is the null map: we denote it by $F(X)$. Yet, given any subset $X$ of
a commutative monoid $M$, the following map\vspace{-0.15cm}
\[
\xymatrix@R-10mm
{
\ar@{}[r]|(0.2){\begin{minipage}{3mm}$\Phi^X_M:$\end{minipage}} & F(X) \ar[r] & M \\
& \phi \ar@{|->}[r] & \underset{x\in X}{\prod}x^{\phi(x)}
}
\]
is a well-defined morphism of monoids. A well-known result asserts
that the following are equivalent \cite{Algebra}:\hiddencom{Citation.} \hiddencom{Au fait, y a-t-il
vraiment des cas où les irréductibles n'engendrent pas l'ensemble ?
Réponse trouvée avec les exemples suivants: cas de $(\Z,+)$ qui n'a
pas d'irréductibles.}
\begin{enumerate}
\item the mapping $\Phi^{I(M)}_M$ is an isomorphism of monoids\medskip
\item the set $I(M)$ generates\footnote{$X\subseteq M$ generates $M$ when
  all its elements can be written as a product of elements of $X$. The
  product of the empty set being defined as the neutral element. Remark
  then that ``$I(M)$ generates $M$'' implies that the only unit of $M$
  is its neutral element.} $M$ and $I(M)=P(M)$\medskip
\item any element of $M$ can be written as a product of irreducible
elements in a unique way up to permutation of terms (unique
decomposition property).
\end{enumerate}
In this case $M$ is said to be a {\bf free} commutative monoid.

\hiddencom{Ce n'est pas un peu bizarre comme endroit pour un bout de related
  works ?}

Two standard examples of free commutative monoids are given by the
set of non zero natural numbers $\N\backslash\{0\}$ together with
multiplication (the unit is $1$ and the irreducible elements are the
prime numbers) and
the set of natural numbers $\N$ together with addition (the unit is
$0$ and the only irreducible element is $1$).

However, neither the multiplicative monoid $\Z\backslash\{0\}$ nor the
additive group $\Z$ are free commutative monoids since they both
contain a non trivial unit, namely $-1$ in both cases.

Also note that all the non zero elements of the additive
monoid $\R_+$ are primes though it does not contain a single
irreducible element.

A more intricate phenomenon arises with polynomials
\cite{TNJHProBir}: the (multiplicative) commutative monoid
$\N[X]\backslash\{0\}$ of non zero polynomials with natural
coefficients is not free. Indeed, although each element of
$\N[X]\backslash\{0\}$ is a product of irreducible polynomials these
decompositions are not unique: we have
$$
(1+X)(1+X^2+X^4)=(1+X^3)(1+X+X^2)
$$
where all of the polynomials $1+X$, $1+X^2+X^4$, $1+X^3$ and $1+X+X^2$
are irreducible (which is not the case in the monoid of polynomials
with coefficients in $\Z$, indeed the ring $\Z[X]$ is {\bf factorial}
\cite{Algebra}).


%
%
\subsection{Cartesian Product and Commutation}

The geometric model of a concurrent program is a set of
  points in an euclidean space of finite dimension. Thus each point
  can be represented by the tuple of its coordinates, and a
  geometric model becomes a set of tuples (of same length which
  corresponds to the dimension of the space).
  The cartesian product on such structures is the following:
  \[
  X\times
  Y\quad=\quad\big\{\ (x_1,...,x_n,y_1,...,y_k)\ \big|\ (x_1,...,x_n)\in
  X,(y_1,...,y_k)\in Y\ \big\}
  \]
  However, this operator is not commutative whereas the parallel
  composition of programs should be so. Thus, in order to model
  parallel composition, we make the operator $\times$ commutative
  monoid through quotient by permutation of coordinates.  In the next
  subsection we prove a freeness theorem for a monoid generalizing
  this idea: tuples of (real) coordinates are replaced by words over
  an arbitrary (potentially infinite) alphabet. 
The geometric model of a PV program therefore belongs to a free
commutative monoid and thus admits a unique decomposition of
irreducible elements, from which the processes factorization is
deduced.

\subsection{Homogeneous Sets of Words\manu{ over an
  alphabet $\A$}}

Let $\A$ be a set called the alphabet. The {\em non commutative}
monoid of words $\A^\ast$ consists on the finite sequences of elements
of $\A$ together with {\bf concatenation}. Given words $w$ and
$w'$ of length $n$ and $n'$, the word $w\ast w'$ of length $n+n'$ is
defined by
$$
(w\ast w')_k=\left\{
\begin{array}{l@{\hspace{4mm}\mbox{if}\hspace{2mm}}l}
w_k & \phantom{n+1}1\leqslant k\leqslant n \\
w'_{k-n} & \phantom{1}n+1\leqslant k\leqslant n+n'
\end{array}
\right.
$$
The length of a word $w$ is also refered to as $\ell(w)$. A {\bf
  subword} of $w$ is a word of the form $w\circ\phi$ where $\phi$ is a
strictly increasing map
$\{1,\ldots,n\}\rightarrow\{1,\ldots,\ell(w)\}$. Hence a subword of
$w$ is also entirely characterized by the image of the increasing map
$\phi$ i.e.\ by a subset of $\{1,\ldots,\ell(w)\}$. If $A$ is the
image of $\phi$ then we write $w\circ A$ instead of $w\circ\phi$.

The $n^{\text{th}}$ {\em symmetric group} $\mathfrak{S}_n$ (the group
of permutations of the set $\{1,...,n\}$) acts on the set of words of
length $n$ by composing on the right, that is for all
$\sigma\in\mathfrak{S}_n$ and all word $w$ of length $n$ we have
$$\sigma\cdot w:=w\circ\sigma=(w_{\sigma(1)}\cdots w_{\sigma(n)})$$

The concatenation extends to sets of words. Given $S,S'\subseteq
\A^\ast$, define
$$
S\ast S':=\{w\ast w'\ |\ w\in S;w'\in S'\}
$$ 
Remark that this concatenation of sets corresponds to the cartesian product.

The set ${\Pmaj}(\A^\ast)$ of subsets of $\A^\ast$ is thus endowed
with a structure of {\em non commutative} monoid whose neutral element
is $\{\epsilon\}$: the singleton containing the empty word. Note that
the empty set $\emptyset$ is the {\em absorbing} element of
${\Pmaj}(\A^\ast)$, that is for all $S\subseteq \A^\ast$ we have
$$
\emptyset\ast S=S\ast\emptyset=\emptyset
$$

A subset $H$ of $\A^\ast$ is said to be {\bf homogeneous} when all the
words it contains share the same length $n$\manu{, which is then
  called the length of $H$ and denoted by $\ell(H)$}. By analogy with
the geometric construction, $n$ is called the dimension of $H$ and
denoted by $\dd(H)$. The symmetric group $\mathfrak{S}_n$ acts on the
set of homogeneous set of dimension $n$ in a natural way by applying
the same permutation to all words:
$$
\sigma\cdot H:=\{\sigma\cdot w\ |\ w\in H\}
$$

The homogeneous subsets of $\A^\ast$ form a sub-monoid
${\Pmaj}_h(\A^\ast)$ of ${\Pmaj}(\A^\ast)$ and can be equipped with
an equivalence relation as follows: write $H\sim H'$ when
$\dd(H)=\dd(H')=n$ and there exists $\sigma\in\mathfrak{S}_n$ such
that $H'=\sigma\cdot H$. Moreover, for two permutations
$\sigma\in\mathfrak{S}_n$ and $\sigma'\in\mathfrak{S}_{n'}$, define the
{\bf juxtaposition} $\sigma\otimes\sigma'\in\mathfrak{S}_{n+n'}$ as:
$$
\sigma\otimes\sigma'(k):=
\left\{
\begin{array}{c@{\hspace{4mm}\mbox{if}\hspace{2mm}}l}
\sigma(k) & \phantom{n+1}1\leqslant k\leqslant n\\
\big(\sigma'(k-n)\big)+n & \phantom{1}n+1\leqslant k\leqslant n+n'
\end{array}
\right.
$$
A Godement-like exchange law is satisfied, which ensures that $\sim$
is actually a congruence:
$$
(\sigma\cdot H)\ast(\sigma'\cdot H')=(\sigma\otimes\sigma')\cdot(H\ast H')
$$
Hence the quotient ${\Pmaj}_h(\A^\ast)/\!\!\sim$ from which the
absorbing element has been removed is still a monoid called the {\bf
  homogeneous monoid} over $\A$ and denoted by ${\Hmaj}(\A)$. Moreover
the homogeneous monoid is commutative and its only unit is the
singleton $\{\epsilon\}$. Remark that if the alphabet $\A$ is a
singleton (resp. the empty set) then the homogeneous monoid
${\Hmaj}(\A)$ is isomorphic to $(\N,+,0)$ (resp. the null monoid).
%
~\medskip

\noindent
\hspace{-0.3cm}
\fbox{
\begin{minipage}{\linewidth}
\begin{theorem}
\label{homogeneous_is_commutative_free}
For any set $\A$ the homogeneous monoid over $\A$ is free.
\end{theorem}
\end{minipage}
}\medskip

\begin{proof}
We check the conditions 1-3 which characterize the free commutative
monoids (see Section 3.2).  \manu{The monoid ${\Hmaj}(\A)$ is
  obviously commutative.}  Since $\dd(H\ast H')=\dd(H)+\dd(H')$ we
deduce from a straightforward induction on the dimension of elements
of ${\Hmaj}(\A)$ that they can all be written as products of
irreducible elements: $I({\Hmaj}(\A))$ generates ${\Hmaj}(\A)$.

Now suppose $H\manu{_0}$ is an irreducible element of
  ${\Hmaj}(\A)$ which divides
$H_1\ast H_2$ and pick $S\manu{_0}$, $S_1$ and $S_2$ respectively from the
equivalence classes $H\manu{_0}$,
$H_1$ and $H_2$. Define $n=\dd(H)$, $n_1=\dd(H_1)$ and
$n_2=\dd(H_2)$, and remark that $n=n_1+n_2$.
There exists $\sigma\in\mathfrak{S}_n$ and some
$S_3$ such that $\sigma\cdot(S_1\ast S_2)=S\ast S_3$ in
${\Pmaj}_h(\A^\ast)$. Suppose in addition that $H$ does not divide
$H_1$ nor $H_2$, then we have $A_1\subseteq\{1,...,n_1\}$ and
$A_2\subseteq\{1,...,n_2\}$ s.t.\ $A_1\not=\emptyset$,
$A_2\not=\emptyset$ and $\sigma(A_1\cup A_2')=\{1,...,n\}$ where
$A_2':=\{a+n_1\ |\ a\in A_2\}$. Then we have a non trivial factoring
$S=S'_1\ast S'_2$ where
$$
S'_1:=\big\{w\circ A_1\ \big|\ w\in S_1  \big\}\text{ and }S'_2:=\big\{w\circ A_2\ \big|\ w\in S_2  \big\}
$$
This contradicts irreducibility of $H$. Hence $H$ divides
  $H_1$ or $H_2$ and thus $H$ is prime.
So any irreducible element of ${\Hmaj}(\A)$ is prime:
$I({\Hmaj}(\A))\subseteq P({\Hmaj}(\A))$.

Finally, suppose $H$ is a prime element of ${\Hmaj}(\A)$ such that
$H=H_1\ast H_2$. In particular $H$ divides $H_1\ast H_2$, and since
$H$ is prime it divides $H_1$ or $H_2$. Both cases being symmetrical,
suppose $H$ divides $H_1$. In particular $\dd(H)\leq\dd(H_1)$. On the
other hand $\dd(H)=\dd(H_1)+\dd(H_2)$, and thus $\dd(H_2)\leq
0$. Dimensions being natural numbers, we deduce that $\dd(H_2)=0$ and
then that $H_2=\{\epsilon\}$. Hence $H$ is irreducible, and
$I({\Hmaj}(\A))=P({\Hmaj}(\A))$.
\end{proof}
One of the worthy feature of the construction is that any binary
relation $\diamond$ over ${\Pmaj}_h(\A)$ which is compatible with the
product and satifies
$$ \forall S,S'\in {\Pmaj}_h(\A)\ \big(\dd(S)=\dd(S')=n\text{ and
}S\diamond
S'\ \Rightarrow\ \forall\sigma\in\mathfrak{S}_n\ (\sigma\cdot
S)\diamond(\sigma\cdot S')\big)
$$ can be extended to a relation on ${\Hmaj}(\A)$ which is still
compatible with the product. Actually it suffices to set $H\diamond
H'$ when $\dd(H)=\dd(H')=n$
and there exists a representative $S$ of
$H$ and a representative $S'$ of $H'$ such that for all
$\sigma\in\mathfrak{S}_n$ we have $(\sigma\cdot S)\diamond(\sigma\cdot
S')$. In addition, if the relation $\diamond$ satisfies
$$ \forall S,S'\in {\Pmaj}_h(\A)\ S\diamond
S'\ \Rightarrow\ \dd(S)=\dd(S')
$$ then the quotient map is compatible with $\diamond$ and its
extension. The relation of inclusion $\subseteq$ over ${\Pmaj}_h(\A)$
obviously satisfies these properties and therefore extends to
${\Hmaj}(\A)$.
\subsection{Cubical Areas}
\label{Cubical_Areas}
A {\bf cube} of dimension $n$ is a word of length $n$ on the alphabet
${\Imaj}$ of {\em non-empty} intervals of $\R$. The elements of
${\Hmaj}({\Imaj})$ are called the {\bf cubical coverings}. Furthermore
the homogeneous monoid ${\Hmaj}({\Imaj})$ is endowed with a preorder
arising from the inclusion on ${\Imaj}$. Indeed, given two sets of
cubes of the same length $S$ and $S'$ we write $S\preccurlyeq S'$ when
for all cubes $C\in S$ there exists a cube $C'\in S'$ such that
$C\subseteq C'$. The relation $\preccurlyeq$ provides the monoid
${\Pmaj}({\Imaj})$ with a preorder that can be extended to
${\Hmaj}({\Imaj})$ by setting
$H\preccurlyeq H'$ when $\dd(H)=\dd(H')=n$ and there exists a
representative $S$ of $H$ and a representative $S'$ of $H'$ such that
for all $\sigma\in\mathfrak{S}_n$ we have $(\sigma\cdot
S)\preccurlyeq(\sigma\cdot S')$.
We now establish a Galois connection between $({\Hmaj}(\R),\subseteq)$
and $({\Hmaj}({\Imaj}),\preccurlyeq)$. Given a cubical covering $F$ we
define $\gamma(F)$ as
$$
\Big\{\bigcup_{C\in S}C\ \Big|\ S\in F\Big\}
$$ 
Furthermore $\gamma$ is a morphism of monoids and if $F\preccurlyeq
F'$ then $\gamma(F)\subseteq\gamma(F')$.

\noindent
Conversely, given some $S$ in ${\Pmaj}_h(\R^\ast)$ the collection of
$n$-dimensional cubes $C$ such that $C\subseteq S$, ordered by
inclusion, is a semilattice whose maximal elements are called the {\bf
  maximal cubes} of $S$. The set $M_S$ of maximal cubes of $S$ is
homogeneous and for all $\sigma\in\mathfrak{S}_n$, \ $\sigma\cdot
M_S=M_{\sigma\cdot S}$. Then given $H\in{\Hmaj}(\R)$ we define
$\alpha(H)$ as 
$$
\Big\{M_S\ \Big|\ S\in H\Big\}
$$ 
Furthermore $\alpha$ is a morphism of monoids and if
$H\subseteq H'$ then $\alpha(H)\subseteq\alpha(H')$. Then we have a
Galois connection: 
~\medskip

\noindent
\hspace{-0.3cm}
\fbox{
\begin{minipage}{\linewidth}
\begin{proposition}
\label{Galois_connection}
$\gamma\circ\alpha=\idt_{{\Hmaj}(\R)}$ and\ \ $\idt_{{\Hmaj}({\Imaj})}\preccurlyeq\alpha\circ\gamma$.
\end{proposition}
\end{minipage}
}\medskip

\noindent
Given $H\in{\Hmaj}(\R)$ and $F\in{\Hmaj}({\Imaj})$ we say that $F$ is
a cubical covering of $H$ when $\gamma(F)=H$. The {\bf cubical areas}
are the elements $H$ of ${\Hmaj}(\R)$ which admit a {\em finite}
cubical covering. The collection of cubical areas (resp. finite
cubical coverings) forms a sub-monoid \texttt{Are} of ${\Hmaj}(\R)$
(resp. \texttt{Cov} of ${\Hmaj}({\Imaj})$). The restrictions of the
morphisms $\gamma$ and $\alpha$ to \texttt{Cov} and \texttt{Are}
induce another Galois connection.~\medskip

\noindent
\hspace{-0.3cm}
\fbox{
\begin{minipage}{\linewidth}
\begin{proposition}
\label{Galois_connection2}
$\gamma\circ\alpha=\idt_{\texttt{Are}}$ and\ \ $\idt_{\texttt{Cov}}\preccurlyeq\alpha\circ\gamma$.
\end{proposition}
\end{minipage}
}\medskip

\noindent
Moreover, the morphisms $\gamma$ and $\alpha$ of Proposition
\ref{Galois_connection2} induce a pair of isomorphisms of commutative
monoids between \texttt{Are} and the collection of fixpoints of
$\alpha\circ\gamma$. A submonoid of a free commutative monoid may not
be free. Yet, under a simple additional hypothesis this pathological
behaviour is no more possible. We say that a submonoid $P$ of a monoid
$M$ is {\bf pure} when for all $x,y\in M$, $x\ast y\in P\ \Rightarrow\ x\in
P$ and $y\in P$.~\medskip

\noindent
\hspace{-0.3cm}
\fbox{
\begin{minipage}{\linewidth}
\begin{lemma}
\label{pure_submonoids}
Every pure submonoid of a free commutative monoid is free. 
\end{lemma}
\end{minipage}
}\medskip
\begin{proof}
Let $P$ be a pure submonoid of a free commutative monoid $M$. Let $p$
be an element of $P$ written as a product $x_1\cdots x_n$ of
irreducible elements of $M$. Each $x_i$ is obviously an irreducible
element of $P$ so any element of $P$ can be written as a product of
irreducible elements of $P$. Furthermore any irreducible element of
$P$ is also an irreducible element of $M$ because $P$ is pure in
$M$. It follows that any elements of $P$ can be written as a product
of irreducible elements of $P$ in a unique way i.e.\ $P$ is free.
\end{proof}

\noindent
Then we have:\medskip
%

\noindent
\hspace{-0.3cm}
\fbox{
\begin{minipage}{\linewidth}
\begin{theorem}
\label{monoid_of_cubical_areas_is_commutative_free}
The commutative monoid of cubical areas is free and has infinitely
many irreducible elements.
\end{theorem}
\end{minipage}
}\medskip
\begin{proof}
Let $X$ and $X'$ be two elements of ${\Hmaj}(\R)$ and suppose $X\ast
X'$ belongs to \texttt{Are}. Since both $\alpha$ and $\gamma$ are
morphisms of monoids we have $\alpha\circ\gamma(X\ast
X')=\alpha\circ\gamma(X)\ast\alpha\circ\gamma(X')$ which is finite. It
follows that both $\alpha\circ\gamma(X)$ and $\alpha\circ\gamma(X')$
are finite. Hence $X$ and $X'$ actually belongs to \texttt{Are}, which
is thus free as a pure submonoid of ${\Hmaj}(\R)$.
\end{proof}
\section{Effective Factoring of Cubical Areas}
Beyond their theoretical usefulness, the maximal cubes provide the
data structure which allows to handle algorithmically cubical areas,
as in the static analyzer \texttt{ALCOOL} which is devoted to the
study of parallel programs.
\label{algorithm}
\subsection{Implementation}
We need an algorithm which performs decompositions in ${\Hmaj}(\A)$,
its implementation is directly based on the proof of the Theorem
\ref{homogeneous_is_commutative_free}: $H\in{\Hmaj}(\A)$ is reducible
if and only if there exists some representative $S$ of $H$ which
admits a non trivial decomposition in ${\Pmaj}_h(\A^\ast)$. In order
to describe the algorithm we define
$$
S\circ A:=\big\{w\circ A \ |\ w\in S\big\}
$$ for any $S\in{\Pmaj}_h(\A^\ast)$ and
$A\subseteq\{1,...,\dd(S)\}$. Moreover for $w'\in\A^\ast$ with
$\ell(w')\leqslant\dd(S)$, and $A^c$
the complement of $A$ (in $\{1,...,\dd(S)\}$), we define the set of
words 
$$
\Psi(w',A,S):=\big\{w\circ A^c\ |\ w\in S\ \text{ and }\ S\circ A=w'\big\}
$$ Then the class $[S\circ A]\in{\Hmaj}(\A)$ divides $H$ if and only
if for all $w'\in S\circ A$ one has $\Psi(w',A,S)=[S\circ A^c]$. In
the monoid ${\Hmaj}(\A)$ we thus have
$$
[S\circ A]\ast[S\circ A^c]=H
$$ Then we look for some divisor of $H$ by testing all the non empty
subsets $A$ of $\{1,\ldots,\dd(S)\}$ according to the following total
ordering
$$
A\leqslant A'\ \text{ when }\ |A|<|A'|\ \text{ or }\ (|A|=|A'|\ \text{ and }\ |A|\ssbeq_{\text{lex}}|A'|)  
$$ where $\ssbeq_{\text{lex}}$ is the lexicographic ordering. Doing
so, we know that if $A$ is the first value such that $[S\circ A]$
divides $H$, then $[S\circ A]$ is irreducible. Moreover we have
$\dd([S\circ A])=|A|$ and for all $H_0,H_1\in{\Hmaj}(\A)$, 
$\dd(H_0\ast H_1)=\dd(H_0)+\dd(H_1)$ hence we can
suppose $$|A|\leqslant\frac{\dd(H)}{2}+1$$

\noindent
The software \texttt{ALCOOL} is entirely written in \texttt{OCaml}. The complexity of
the decomposition algorithm implemented in it is exponential in the
dimension $n$ of the cubical area since it checks all the subsets of
$\{0,\ldots,n-1\}$. However the computation time actually devoted to
the decomposition is rather small with regard to the global execution
time required by the whole analysis. Indeed the algorithm which builds
the state space of the program, though it has the same theoretical
complexity as the decomposition algorithm, has to handle heavier
structures.
\subsection{A detailed example}
We treat the case of the program $\Sigma$ given in
Example~\ref{IndExample}. Its
geometric model is given on the left hand side of
Figure~\ref{CubExample}. Applying the permutation $(2,3)$ we obtain
the right hand side set.
\begin{figure}
  \caption{Cubical area of Example~\ref{IndExample}}
\label{CubExample}
\begin{center}
\begin{minipage}{0.4\textwidth}
\begin{verbatim}
   [0,1[*[0,1[*[0,-[*[0,-[
|| [0,1[*[4,-[*[0,-[*[0,-[
|| [0,1[*[0,-[*[0,-[*[0,1[
|| [0,1[*[0,-[*[0,-[*[4,-[
|| [4,-[*[0,1[*[0,-[*[0,-[
|| [4,-[*[4,-[*[0,-[*[0,-[
|| [4,-[*[0,-[*[0,-[*[0,1[
|| [4,-[*[0,-[*[0,-[*[4,-[
|| [0,-[*[0,1[*[0,1[*[0,-[
|| [0,-[*[0,1[*[4,-[*[0,-[
|| [0,-[*[0,-[*[0,1[*[0,1[
|| [0,-[*[0,-[*[0,1[*[4,-[
|| [0,-[*[4,-[*[0,1[*[0,-[
|| [0,-[*[4,-[*[4,-[*[0,-[
|| [0,-[*[0,-[*[4,-[*[0,1[
|| [0,-[*[0,-[*[4,-[*[4,-[
\end{verbatim}
\end{minipage}
\hskip0.1\linewidth
\begin{minipage}{0.4\textwidth}
\begin{verbatim}
   [0,1[*[0,-[*[0,1[*[0,-[
|| [0,1[*[0,-[*[4,-[*[0,-[
|| [0,1[*[0,-[*[0,-[*[0,1[
|| [0,1[*[0,-[*[0,-[*[4,-[
|| [4,-[*[0,-[*[0,1[*[0,-[
|| [4,-[*[0,-[*[4,-[*[0,-[
|| [4,-[*[0,-[*[0,-[*[0,1[
|| [4,-[*[0,-[*[0,-[*[4,-[
|| [0,-[*[0,1[*[0,1[*[0,-[
|| [0,-[*[4,-[*[0,1[*[0,-[
|| [0,-[*[0,1[*[0,-[*[0,1[
|| [0,-[*[0,1[*[0,-[*[4,-[
|| [0,-[*[0,1[*[4,-[*[0,-[
|| [0,-[*[4,-[*[4,-[*[0,-[
|| [0,-[*[4,-[*[0,-[*[0,1[
|| [0,-[*[4,-[*[0,-[*[4,-[
\end{verbatim}  
\end{minipage}
\end{center}
\end{figure}

\noindent
Then we can check that the (right hand side of
Figure~\ref{CubExample}) cubical area can be written as
\begin{center}
$\Big($\texttt{[0,1[*[0,-[ $\|$ [4,-[*[0,-[ $\|$ [0,-[*[0,1[ $\|$ [0,-[*[4,-[}$\Big)^2$
\end{center}
Then we have 
$$
(2,3)\cdot\big\{\{1,2\},\{3,4\}\big\}=\big\{\{1,3\},\{2,4\}\big\}
$$ and it follows that in the program $\Sigma$ the sets of processes
$\{\pi_1,\pi_3\}$ and $\{\pi_2,\pi_4\}$ run independently from each
other.
%
%
\subsection{Benchmarks}
We describe some programs upon which the algorithm has been tested. The
program $\Sigma_{n_1,\ldots,n_k}$ is made of $k$ groups of processes:
for all $i\in\{1,...,k\}$ it contains $n_i$ copies of the
process $$P(a_i).P(b).V(b).V(a_i)$$
where $a_i$ is a mutex and $b$ is a semaphore of arity $k+1$.
All processes then share the resource $b$, but as for $\Sigma$ in
Example~\ref{IndExample} the $k$ groups are actually independent.
On the other hand the program $\Sigma'_{n_1,\ldots,n_k}$ is the same
as $\Sigma_{n_1,\ldots,n_k}$ but with $b$ of arity only $k$, which forbids
any decomposition. The $n$-philosophers programs implement the
standard $n$ dining philosophers algorithm.

The benchmark table of Figure~\ref{Bench} has been obtained using the Unix command
\texttt{time} which is not accurate. Hence these results have to be
understood as an overapproximation of the mean execution time.
\begin{figure}
\caption{Benchmarks}
\label{Bench}
\begin{center}
\begin{tabular}{|c|c|c|}
\hline
Example & Time (in sec.) & Decomp. \\
\hline
6 philosophers & $0.2$ & No\\
\hline
7 philosophers & $0.7$ & No\\
\hline
8 philosophers & $3.5$ & No\\
\hline
9 philosophers & $21$ & No\\
\hline
10 philosophers & $152$ & No\\
\hline
\end{tabular}\vskip2mm
\begin{tabular}{|c|c|c||c|c|c|}
\hline
Example & Time (in sec.) & Decomp. & Example & Time (in sec.) & Decomp. \\
\hline\hline
$\Sigma_{2,2}$ & $0.1$ & $\{1,3\}\{2,4\}$ & $\Sigma'_{2,2}$ & $0.1$ & No\\
\hline\hline
$\Sigma_{2,2,2}$ & $0.1$ & $\{1,4\}\{2,5\}\{3,6\}$ & $\Sigma'_{2,2,2}$ & $0.3$ & No \\
\hline
$\Sigma_{3,3}$ & $0.13$ & $\{1,3,5\}\{2,4,6\}$ & $\Sigma'_{3,3}$ & $0.52$ & No \\
\hline\hline
$\Sigma_{2,2,2,2}$ & $0.13$ & $\{1,5\}\{2,6\}\{3,7\}\{4,8\}$ & $\Sigma'_{2,2,2,2}$ & $7.1$ & No \\
\hline
$\Sigma_{4,4}$ & $1$ & $\{1,3,5,7\}\{2,4,6,8\}$ & $\Sigma'_{4,4}$ & $33$ & No\\
\hline\hline
$\Sigma_{3,3,3}$ & $1.5$ & $\{1,4,7\}\{2,5,8\}\{3,6,9\}$ & $\Sigma'_{3,3,3}$ & $293$ & No \\
\hline
$\Sigma_{4,5}$ & $6.1$ & $\{1,3,5,7\}\{2,4,6,8\}$ & $\Sigma'_{4,5}$ & $327$ & No\\
\hline\hline
$\Sigma_{5,5}$ & $50$ & $\{1,3,5,7,9\}\{2,4,6,8,10\}$ & $\Sigma'_{5,5}$ & $2875$ & No\\
\hline
\end{tabular}
\end{center}
\end{figure}
\manu{Yet we can observe
that the execution time explodes when analyzing a $\Sigma'$-like
program instead of a $\Sigma$-like one. The decomposition algorithm is
not to blame, indeed in these examples, it seems more than $95\%$ of
the time required to analyze $\Sigma'_{3,3,3}$ is devoted to the
calculation of the geometric semantics.}
It is also worth remarking that our algorithm is efficient when the
cubical area to decompose is actually a cartesian product of several
irreducible cubical areas of small dimension. This remark should be
compared with the fact that the standard decomposition algorithm of
integer into primes is very efficient on products of small prime
numbers.
\section{Conclusion}
\noindent{\bf Related work.}

The problem of decomposition of concurrent programs in {\em CCS}-style
has been studied in \cite{JFGFMDecomposition}
and~\cite{RMFMDecomposition}. By the possibility of using semaphores
of arbitrary arity, our work seems to go beyond this previous
approach. Also note that the silent and synchronous communication
mechanism of {\em CCS} can be given a straightforward geometric
interpretation which falls in the scope of the present
discussion. However, the link between bisimilarity in {\em CCS} and
isomorphic geometric interpretations is still to be explored to make
clear the relations between these works.

In~\cite{BLVvODecomposition} {\em B. Luttik} and {\em V. van Oostrom}
have characterized the commutative monoids with unique decomposition
property as those which can be provided with a so-called decomposition
order. In the case where the property holds, the divisibility order always
fits. Yet, there might exist a more convenient one. Unfortunately, in
the current setting the authors are not aware of any such order
yielding direct proofs. Nevertheless it is worth noticing that this
approach is actually applied for decomposition of processes in a
normed ACP theory for which a convenient decomposition order exists.\medskip

\noindent{\bf Conclusion.}

This paper uses a geometric semantics for concurrent programs, and
presents a proof of a unique decomposition property together with an
algorithm working at this semantic level
(Theorem~\ref{monoid_of_cubical_areas_is_commutative_free}). 
The main strength of this work is that it applies to any concurrent
program yielding a cubical area. Example of features allowed in this
setting are: semaphores, synchronisation barriers,
synchronous as well as asynchronous communications (with finite or
infinite message queues), conditional branchings. In fact we can even
consider loops provided we replace the set ${\Imaj}$ of intervals of
the real line $\R$ by the set $\Amaj$ of arcs of the circle.\medskip

\noindent{\bf Future work.}

Actually, a cubical area naturally enjoys a pospace\footnote{shorthand
  for ``partially ordered spaces'' \cite{Nachbin}.}
structure. Pospaces are among simplest objects studied in {\em
  Directed Algebraic Topology}. In particular, a cubical area is
associated with its category of components
\cite{FGHRCompCat,EGEHPraAppGeo,EHCatComCatSanBou} and
\cite{EGEHCompCat2}, which is proven to be finite,
loop-free\footnote{Loop-free categories were introuced in
  \cite{HAOrbihedra,HAExtCompGroups} as ``small categories without
  loop'' or ``scwols''.} and in most cases connected. Then, as the
cubical areas do, these categories together with cartesian product
form a free commutative monoid. It is worth noticing this is actually
the generalization of a result concerning finite posets which has been
established in the early fifties \cite{JHOnProDecPoSet}. Therefore a
program $\Pi$ can be decomposed by lifting the decomposition of the
category of components of its geometric model
$\llbracket\Pi\rrbracket$. In general, the relation between the
decomposition of a cubical area and the one of its category of
components is a theoretical issue the authors wish to investigate.

Another important concern is a clarification of the control constructs
compatible with cubical areas: replacing in some dimensions the
intervals of the real line by the arcs of the circle as mentioned
above corresponds to a global loop, but some richer structures may be
useful.

A final point of interest is the investigation of the exact relation
between our semantic results and the syntactic ones
of~\cite{JFGFMDecomposition,RMFMDecomposition,BLVvODecomposition}.
Indeed they use $CCS$-like syntaxes to describe some classes of
edge-labelled graphs modulo bisimilarity, whereas the category of
components of our models correspond to some other graphs modulo
directed homotopy. Hence the question: what is in this setting the
relation between bisimilarity and homotopy?





\end{document}